\documentstyle[aps]{revtex}
\begin{document}
\title{The Jahn-Teller theorem for the 3d magnetic ions*}
\author{Z. Ropka}
\address{Center for Solid State Physics, \'{s}w. Filip 5, 31-150 Krak\'{o}w.}
\author{R.J. Radwa\'{n}ski}
\address{Center for Solid State Physics, \'{s}w. Filip 5, 31-150 Krak\'{o}w,\\
Inst. of Physics, Pedagogical University, 30-084 Krak\'{o}w, Poland.\\
email: sfradwan@cyf-kr.edu.pl}
\maketitle

\begin{abstract}
We argue that the Jahn-Teller theorem, if applied to 3d-ion compounds, has
to be considered in the orbital+spin space. It is despite of the weakness of
the intraatomic spin-orbit coupling.



Receipt date by Phys.Rev.Lett.: 16.02.1999
\end{abstract}

\date{(16 February 1999)}

There is ongoing strong discussion$^a$ in Phys.Rev.Lett. and Phys.Rev.B
about a relationship between the electronic structure, magnetism, the local
3d-ion surrounding and a (off-cubic) lattice distortion in 3d ionic
compounds like La$_{1-x}$Ca$_x$MnO$_3$, LaCoO$_3$. In this discussion, terms
like the Jahn-Teller (J-T) effect and the J-T distortions appear that
obviously are related with the J-T theorem. This long-time discussion
reveals that the physical picture for 3d-ion compounds is far from a
consensus. On other side, the x-ray-absorption fine structure (XAFS)
experiments provide more and more accurate data on the local surrounding in
the atomic scale.

The aim of this short Letter is to put attention that the degeneracy space
for the consideration of the J-T theorem for the 3d ions is the orbital+spin
space. It is in contrast to the current-literature that considers the J-T
theorem, if applied to 3d ions, in the orbital space only.

In 1937 Jahn and Teller had pointed out$^{1,2}$ that no non-linear molecule
could be stable in a degenerate state. It means that one would expect a
spontaneous distortion to undergo in order to remove this degeneracy. In the
current-literature the J-T theorem and related with it the degeneracy, if
applied to 3d ions, is considered in the orbital space. It likely repeats
the text-book knowledge (Ref.2 p.193 line 4 bottom, Ref. 3 p. 212. line 12
top, Ref. 4 p.40 line 8 bottom). This way of thinking is commonly used for
the description of the 3d ions and 3d-ion compounds basing on the weakness
of the spin-orbit coupling for these ions.

We argue that the J-T theorem has to be considered in the orbital+spin
space. It is the consequence of the spin-orbit coupling that, though weak in
the 3d ions, always exists.

The demanded orbital+spin space for the consideration of the J-T theorem is
consistent with the standard treatment of rare-earth-ion compounds.

In conclusion, we argue that the J-T theorem, if applied to 3d-ion
compounds, has to be considered in the orbital+spin space.

*This paper has been submitted 16.02.1999 to Phys.Rev.Lett. [LB7427] as a
consequence of a erroneous report of the referee of Phys.Rev.Lett. on our
paper [5] that ''the Jahn-Teller theorem only applies to orbitally
degenerate levels and therefore does not apply to the system considered in
the paper'' [LL6530, report A from 4.12.1997], i.e. the d$^8$ system in the
octahedral crystal field. This report was taken by The Editor of
Phys.Rev.Lett. Dr G.Wells as the base for the rejection of the a/m paper. In
the paper LL6530 we have shown [5] that the d$^8$ system (the Ni$^{2+}$ ion)
in the octahedral surrounding is the Jahn-Teller ion, despite of the orbital
singlet ground state. The long discussion on this paper and others (e.g.
Ref. 6, submitted 30.05.1997 as LE6925) has resulted in the undertaking by
the Managing Editor (Dr G.Wells) and the Editor in Chief (Dr M.Blume,
30.03.2000) of a special discriminating policy with respect to my papers.
Such the policy is the manipulation of Science by the Editors of
Phys.Rev.Lett. and violates the fundamental scientific rules. Our request to
publish our paper with negative referee reports has been ignored.

$^a$In the submitted letter of 15.02.1999 some examplanary papers discussing
the J-T effect have been mentioned: Phys.Rev.Lett. 80 (1998) 853,
Phys.Rev.Lett. 76 (1996) 4825, Phys.Rev.Lett.77 (1996) 5296.

\end{document}